\documentclass[twocolumn,seceq]{jpsj2}

\def\runauthor{Tatsuya Nagao and Jun-ichi Igarashi}

\title{
Resonant X-Ray Scattering on the $M$-Edge Spectra
from Triple-\textbf{k} Structure Phase in U$_{0.75}$Np$_{0.25}$O$_{2}$
and UO$_{2}$
}

\author{Tatsuya Nagao\thanks{E-mail address: tnagao@phys.sci.gunma-u.ac.jp}
and Jun-ichi Igarashi$^{1}$}

\inst{Faculty of  Engineering, Gunma University, Kiryu, Gunma 376-8515 \\
$^{1}$Faculty of Science, Ibaraki University, Mito, Ibaraki 310-8512}

\recdate{\today}

\abst{
We derive an expression for the scattering amplitude of 
resonant x-ray scattering under the assumption that the
Hamiltonian describing the intermediate state preserves
spherical symmetry.
On the basis of this expression,
we demonstrate that the energy profile of the RXS spectra 
expected near U and Np $M_4$ edges
from the triple-\textbf{k} antiferromagnetic
ordering phase in UO$_{2}$ and U$_{0.75}$Np$_{0.25}$O$_{2}$
agree well with those from the experiments.
We demonstrate that the spectra in the $\sigma$-$\sigma'$ and 
$\sigma$-$\pi'$ channels exhibit quadrupole and dipole natures,
respectively.
}

\kword{resonant x-ray scattering, triple-\textbf{k} antiferromagnetic 
ordering, UO$_{2}$, NpO$_{2}$, U$_{0.75}$Np$_{0.25}$O$_2$}

\begin{document}
\maketitle

\section{\label{sec:1}Introduction}

Resonant x-ray scattering (RXS) data obtained from actinide compound
around the $M_{4,5}$ edges supply invaluable information of the
properties of the $5f$ state,\cite{Mannix99,Paixao02,Isaacs90,Bernhoeft03}
since the scattering process consists of
$3d \rightarrow 5f$ transition in the dipolar ($E1$) process.
Recently, we have derived an useful expression for the scattering
amplitude of RXS under the assumption that 
the Hamiltonian describing the intermediate state preserves
the rotational invariance.\cite{Nagao05.2}
Our theory explicitly treats
the energy dependence of the scattering amplitude 
in contrast with the previous theory employing the fast
collision approximation.\cite{Hannon88,Luo93,Lovesey96}
This expression is particularly convenient for
analyzing the energy profile of the RXS spectra from 
actinide $M_{4,5}$ edges such as
NpO$_{2}$, URu$_{2}$Si$_{2}$ and so on.\cite{Nagao05.1,Nagao05.2}
In this paper, we apply the result to investigate the
RXS spectra near U and Np $M_{4}$ edges from the ordered phase
in UO$_{2}$ and U$_{0.75}$Np$_{0.25}$O$_2$.\cite{Wilkins04,Wilkins05}
Analysis of the RXS spectra in these compounds
suggested to show the complicated ordering patterns
such as the triple-${\textbf k}$ type ones
gives useful information complementary
to that from the neutron scattering experiment.

Known actinide dioxides exhibit the cubic CaF$_{2}$ crystal 
structure ($F_{m\overline{3}m}$) at room temperature.
Experiments have confirmed that 
both UO$_{2}$ and U$_{0.75}$Np$_{0.25}$O$_2$
behave as antiferromagnets (AFM) below $T_{\textrm N} =30.8$ K and
$\approx 17$ K, respectively, having transverse triple-\textbf{k} structure
with modulation vector 
$\textbf{Q}=(001)$.\cite{Burlet86,Caciuffo99,Wilkins04}
On the other hand,
recent RXS experiments have suggested longitudinal triple-\textbf{k}
antiferrooctupolar (AFO) ordering phase is plausible in NpO$_{2}$ 
below $T_0=25.5$ K with the same modulation 
vector.\cite{Paixao02,Caciuffo03,Tokunaga05,Kubo05,Kiss03}
In our previous work\cite{Nagao05.2} 
we have investigated the RXS spectra
expected from the longitudinal type of the triple-\textbf{k} AFO phase
in NpO$_{2}$ and obtained fairly good agreement 
with the experiments.
As we shall show, two types of the energy profile of the spectra
($\alpha_1(\omega)$ and $\alpha_2(\omega)$ in Eq. (\ref{eq.amplitude}))
are expected from the triple-\textbf{k} AFM ordering phase,
while only one type of that ($\alpha_2(\omega)$) emerges
from the triple-\textbf{k} AFO ordering phase.
In this work, we concentrate on the former case paying attention
to the difference between the spectra at U and Np sites,
namely, between the $f^2$ and $f^3$ configurations.

\section{\label{sec:2}RXS Amplitude}
Since we are interested in the $5f$ electron systems in which
a localized picture gives reasonable description, we suppose
the initial state at the site $j$ is spanned by
the states within ground multiplet $J$ as
$| 0 \rangle_j$ $=$ $\sum_{m=-J}^{J} c_j(m) | J, m \rangle$.
An incident photon with energy $\hbar \omega$, wave vector $\textbf{k}$,
and polarization $\mbox{\boldmath{$\epsilon$}}$ excites
an electron in core $3d$ level into the unoccupied $5f$ state,
then the excited electron falls into the core level emitting
a scattered photon with energy $\hbar \omega$, wave vector $\textbf{k}'$,
and polarization $\mbox{\boldmath{$\epsilon$}}'$ in the $E1$ process
around the actinide $M_{4,5}$ edges.

When the Hamiltonian describing the
intermediate state of the scattering process
preserves the spherical symmetry, the expression
of the RXS amplitude reduces to an extremely simple form.\cite{Nagao05.2}
This condition is justified when the crystal electric field (CEF) Hamiltonian
and the inter-site interaction are negligible compared with the
energy scale of the intermediate state.
Then, the RXS amplitude at the site $j$ is summarized as\cite{Nagao05.2}
\begin{eqnarray}
 M_j(\mbox{\boldmath{$\epsilon$}}',\mbox{\boldmath{$\epsilon$}},\omega) 
 &=& \alpha_0(\omega)\mbox{\boldmath{$\epsilon$}}' \cdot
                     \mbox{\boldmath{$\epsilon$}} 
 - i\alpha_1(\omega)(\mbox{\boldmath{$\epsilon$}}' 
    \times \mbox{\boldmath{$\epsilon$}})
    \cdot\langle 0|{\bf J}|0\rangle \nonumber\\
 &+& \alpha_2(\omega)\sum_\nu 
 P_{\nu}(\mbox{\boldmath{$\epsilon$}}',\mbox{\boldmath{$\epsilon$}})
 \langle 0|z_{\nu}|0\rangle,
\label{eq.amplitude}
\end{eqnarray}
where quadrupole operators are represented as
$z_1$ $\equiv$ $Q_{x^2-y^2}$ $=$ $(\sqrt{3}/2)$ $(J_x^2-J_y^2)$,
$z_2$ $\equiv$ $Q_{3z^2-r^2}$ $=$ $(3J_z^2-J(J+1))/2$,
$z_3$ $\equiv$ $Q_{yz}$ $=$ $(\sqrt{3}/2)$ $(J_yJ_z+J_zJ_y)$,
$z_4$ $\equiv$ $Q_{zx}$ $=$ $(\sqrt{3}/2)$ $(J_zJ_x+J_xJ_z)$,
and $z_5$ $\equiv$ $Q_{xy}$ $=$ $(\sqrt{3}/2)$ $(J_xJ_y+J_yJ_x)$.
The geometrical factors are defined as
$P_1(\mbox{\boldmath{$\epsilon$}}',\mbox{\boldmath{$\epsilon$}})$
$=$ $(\sqrt{3}/2)(\epsilon'_x\epsilon_x$ $-$ $\epsilon'_y\epsilon_y)$,
$P_2(\mbox{\boldmath{$\epsilon$}}',\mbox{\boldmath{$\epsilon$}})$
$=$ $(2\epsilon'_z\epsilon_z$
$-$ $\epsilon'_x\epsilon_x$ $-$ $\epsilon'_y\epsilon_y)/2$,
$P_3(\mbox{\boldmath{$\epsilon$}}',\mbox{\boldmath{$\epsilon$}})$
$=$ $(\sqrt{3}/2)(\epsilon'_y\epsilon_z$ $+$ $\epsilon'_z\epsilon_y)$,
$P_4(\mbox{\boldmath{$\epsilon$}}',\mbox{\boldmath{$\epsilon$}})$
$=$ $(\sqrt{3}/2)(\epsilon'_z\epsilon_x$ $+$ $\epsilon'_x\epsilon_z)$,
and $P_5(\mbox{\boldmath{$\epsilon$}}',\mbox{\boldmath{$\epsilon$}})$
$=$ $(\sqrt{3}/2)(\epsilon'_x\epsilon_y$ $+$ $\epsilon'_y\epsilon_x)$.
Here, we have suppressed the dependence on the site $j$ in the right hand
side of Eq. (\ref{eq.amplitude}).
Three independent energy profiles $\alpha_{0}(\omega), \alpha_{1}(\omega)$
and $\alpha_{2}(\omega)$ characterize the scalar, dipole and quadrupole
natures, respectively, whose explicit forms have been 
derived in Ref. 5.  
~
\section{\label{sec:3}Application to UO$_{2}$ and U$_{0.75}$Np$_{0.25}$O$_{2}$}

\begin{figure}[t]
\begin{center}
\includegraphics[width=7.6cm]{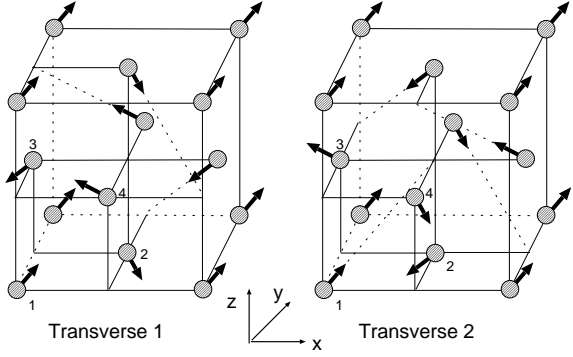}%
\end{center}
\caption{\label{Fig.triplek}  Two types of the
transverse triple-\textbf{k} AFM ordering patterns
corresponding to a modulation vector ${\textbf Q}=(0 0 1)$.
Arrows means vector $(\langle J_x \rangle, \langle J_y \rangle,
\langle J_z \rangle)$ and number 1, 2, 3, and 4 specify the sublattices.
Oxygen ions are omitted.}
\end{figure}

The ordering pattern at U and Np sites in UO$_{2}$ and 
U$_{0.75}$Np$_{0.25}$O$_{2}$ is believed to be the 
transverse triple-\textbf{k} AFM structure (Fig. \ref{Fig.triplek}).
The triple-\textbf{k} structure is defined by all three members
of the star of $\textbf{k}=\langle 001 \rangle$ simultaneously
present on each site.
The order parameter vector of the transverse ordering is composed of three
transverse waves with different \textbf{k}'s.
There are four sublattices 1, 2, 3 and 4 at $(0 0 0)$, 
$(\frac{1}{2} \frac{1}{2} 0)$, $(0 \frac{1}{2} \frac{1}{2})$ and 
$(\frac{1}{2} 0 \frac{1}{2})$, respectively. 
Assuming this ordering pattern is materialized, we investigate
the energy profile and the azimuthal angle dependence of the
RXS spectra at U and Np $M_{4}$ edges.
Since localized picture provides a good stating point,
we construct the ground state of the U ion within the $J=4$ in the
$f^2$ configuration and Np ion within the
$J=\frac{9}{2}$ multiplet in the $f^3$ configuration.  
Note that due to the strong spin-orbit interaction, these ground multiplets
are slightly different from $^{3}H_{4}$ and $^{4}I_{\frac{9}{2}}$ 
terms, respectively, inferred from the perfect LS coupling.

First, we construct the ground state at U site.
Under the influence of the CEF Hamiltonian with cubic symmetry, 
the ground multiplet $J=4$ at U site
splits into one singlet $\Gamma_1$, one doublet $\Gamma_3$
and two triplets $\Gamma_4$ and $\Gamma_5$.
The lowest levels are made of $\Gamma_{5}$.
Because the values of the CEF parameters and the expansion  parameters
of the $\Gamma_{5}$ triplet are irrelevant to the following discussion,
we relegate the actual expressions of them to the 
literature.\cite{Amoretti89}
By introducing the dipole operators,
\begin{equation}
 J_{p} = \left\{ \begin{array}{lcl}
\frac{1}{\sqrt{3}} \left( J_{x}+J_{y}+J_{z} \right) 
     & \textrm{for} & p=111, \\
\frac{1}{\sqrt{3}} \left( J_{x}-J_{y}-J_{z} \right) 
     & \textrm{for} & p=1 \overline{11}, \\
\frac{1}{\sqrt{3}} \left(-J_{x}+J_{y}-J_{z} \right) 
     & \textrm{for} & p=\overline{1}1 \overline{1}, \\
\frac{1}{\sqrt{3}} \left(-J_{x}-J_{y}+J_{z} \right) 
     & \textrm{for} & p=\overline{11}1, \\
\end{array} \right.  \label{eq.tripleJ}
\end{equation}
we can construct the triple-\textbf{k} AFM ordering state 
within the $\Gamma_{5}$ triplet by assigning one of the
eigenstates of $J_p$'s to each sublattice.
Each $J_p$ takes eigenvalues 
$\pm j_{0} \equiv \pm \frac{5}{2}$ and $0$.
Two types of the relevant transverse ordering patterns are
constructed by assigning either one of the eigenstates with 
eigenvalues $\pm \frac{5}{2}$
of $J_{111}, J_{1 \overline{11}}, J_{\overline{1} 1\overline{1}}$
and $J_{\overline{11}1}$ to sublattices 1, 2, 3 and 4 and to
1, 4, 2 and 3, respectively (Fig. \ref{Fig.triplek}). 
For each sublattice assigned operator $J_p$,
the direction of the vector 
$(\langle J_x \rangle, \langle J_y \rangle, \langle J_z \rangle)$,
in which the expectation values are evaluated by the eigenstate of $J_p$,
coincides with that of $p$.

Note that each dipole operator $J_{p}$ commutes with
the corresponding quadrupole operator $Q_p$, which is defined by 
replacing $J_x, J_y$ and $J_z$ in $J_p$ with 
$Q_{yz}, Q_{zx}$ and $Q_{xy}$, respectively.
Each $Q_p$ could be represented as
\begin{equation}
Q_p =\frac{13}{4}\left( \begin{array}{ccc}
 1 & 0 & 0 \\
 0 & 1 & 0 \\
 0 & 0 &-2 \\
 \end{array} \right),
\end{equation}
where the basis are taken by the eigenstates of $J_p$.
It is obvious that the eigenstate of $J_p$ induces 
the quadrupole moment simultaneously. The eigenstates of $J_p$'s
with eigenvalues $\pm j_0$ and $0$ are those of $Q_{p}$'s  
with eigenvalues $\frac{13}{4}$ and $- \frac{13}{2}$, respectively.

Similarly, 
the ground multiplet $J=\frac{9}{2}$ at Np site
splits into one doublet $\Gamma_6$ and two quartets $\Gamma_8^{(1)}$
and $\Gamma_{8}^{(2)}$ due to the CEF Hamiltonian.
The lowest levels are made of $\Gamma_{8}^{(2)}$.
Each $J_p$ within this subspace
takes eigenvalues $\pm j_1$ and $ \pm j_2 (|j_1| > |j_2|)$.
Since the commutation relation $[J_p,Q_p]=0$ also holds,
the eigenstate of $J_p$ induces 
the quadrupole moment simultaneously. The eigenstates of $J_p$'s
with eigenvalues $\pm j_1$ and $\pm j_2$ are those of $Q_{p}$'s  
with eigenvalues $- q_1$ and $+ q_1$, respectively.
Details for the AFM ordering at Np site are found 
in Ref. 5. 

Assuming the eigenstates of $J_p$'s with eigenvalue $-j_0$ 
(or $+j_0$) at U site and those with $-j_1$ (or $+j_1$) at
Np site are realized at each sublattice, 
we calculate the RXS spectra with $\textbf{G}=(1 1 2)$.
Due to the fact that $[J_p,Q_p]=0$, the initial state
takes non-zero expectation values of dipole and quadrupole moment.
This immediately leads to a possibility 
that both two energy profiles $\alpha_1(\omega)$
and $\alpha_2(\omega)$ emerge. 
This is one of the prominent differences when we have investigated 
the RXS spectra from the triple-\textbf{k} AFO state in pure NpO$_{2}$
where only $\alpha_2(\omega)$ is allowed.
For the scattering vector $\textbf{G}=(h h \ell)$ with $h+\ell=odd$,
the interference between $\alpha_{1}(\omega)$ and $\alpha_2(\omega)$
is expected in the $\sigma$-$\pi'$ 
(rotated) channel while only the contribution from 
$\alpha_2(\omega)$ is detected owing to the geometrical factor 
in the $\sigma$-$\sigma'$ (non-rotated) channel.

\begin{figure}
\begin{center}
\includegraphics[width=8.0cm]{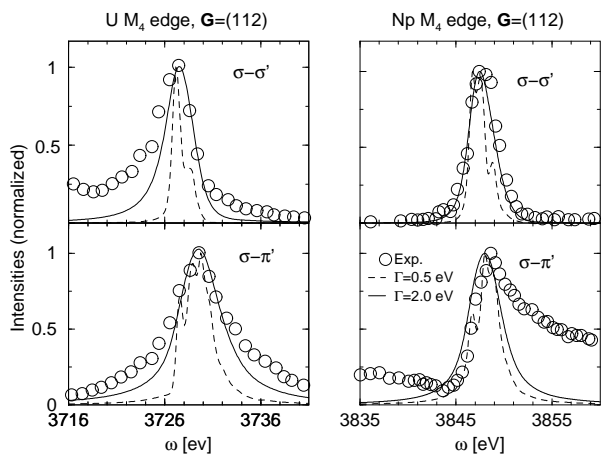}%
\end{center}
\caption{\label{Fig.gam2.0} The RXS spectra as a function of
the photon energy at $\textbf{G}=(112)$ around U and Np $M_4$ edges.
Solid and broken lines are calculated from the transverse triple-\textbf{k} 
AFM phases at U (left) and Np (right) sites with $\Gamma=2.0$ and 0.5 eV,
respectively.
The upper and lower panels are the spectra in the
$\sigma$-$\sigma'$ and $\sigma$-$\pi'$ channels, respectively.
The experimental data are taken from Wilkins \textit{et al}. 
(open circles).\cite{Wilkins04}
All the peak heights are normalized.}
\end{figure}

In order to evaluate the energy profiles $\alpha_1(\omega)$ and
$\alpha_2(\omega)$, we take the intra-atomic Coulomb
interaction and the spin-orbit interaction into full account in 
the preparation of the intermediate state.
Details are described in Ref. 5. 
The value of $\Gamma$, 
the life-time broadening width of the core hole,
is taken as 2 eV (HWHM). 
We display the numerical results in Fig. \ref{Fig.gam2.0}.
We adjusted the core hole energy such that the peak is located at the
experimental position in the $\sigma$-$\sigma'$ channel.
The spectra agree semi-quantitatively well with the experiments.
Since we find $|\alpha_1(\omega)|^2$ is
much larger than $|\alpha_2(\omega)|^2$, 
for instance, $|\alpha_1(\omega)|^2 \sim 125
|\alpha_2(\omega)|^2$ at the U site 
and $\sim 192 |\alpha_2(\omega)|^2$ at the Np site when $\Gamma=2$ eV, 
the interference effect expected 
in the $\sigma$-$\pi'$ channel is negligible.
In experimental data,
the peak position in the $\sigma$-$\pi'$ channel is higher than that
in the $\sigma$-$\sigma'$ channel about 2.0 eV at U site and 0.3 eV at 
Np site. Our results, 1.3 eV and 0.27 eV, respectively, capture such
tendency.
The difference between the spectral shape at U site and that at Np site
is not prominent due to the large value of $\Gamma=2$ eV.
When $\Gamma$ is reduced around 0.5 eV, the spectra show multi-peak 
structure which could distinguish those at U from Np sites when
the drastic improvement of the absorption correction technique would 
been achieved.

\begin{figure}[t]
\begin{center}
\includegraphics[width=7.50cm]{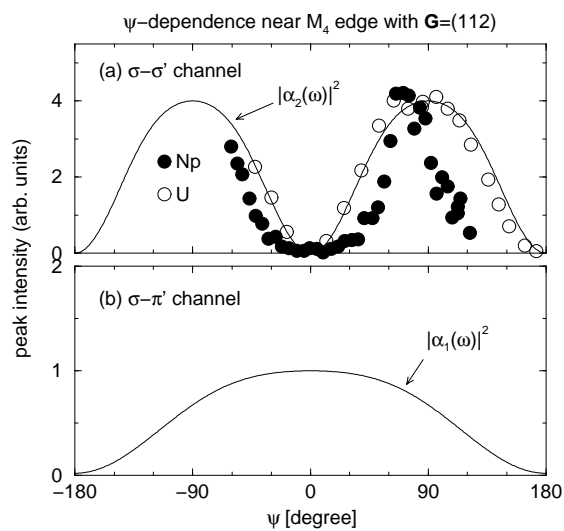}%
\end{center}
\caption{\label{Fig.azim} Azimuthal angle ($\psi$) dependence of the
peak intensity near U and Np $M_{4}$ edges at ${\bf G}=(1 1 2)$
in the transverse triple-{\textbf k} AFM phase.
(a) $\psi$ dependence of $|\alpha_2(\omega)|^2$ in the 
$\sigma$-$\sigma'$ channel. 
Open and filled circles are experimental data
near U $M_4$ edge in UO$_{2}$ (Ref. 11) and 
Np $M_4$ edge in U$_{0.75}$Np$_{0.25}$O$_{2}$ (Ref. 10), 
respectively.
(b) $\psi$ dependence of $|\alpha_1(\omega)|^2$
in the $\sigma$-$\pi'$ channel. 
}
\end{figure}

Fig. \ref{Fig.azim} shows the azimuthal angle dependence of the peak
intensity expected from the transverse triple-\textbf{k} AFM phase
with $\textbf{G}=(112)$.  The spectra are evaluated by incoherent addition 
of the contributions from two transverse ordering patterns.
Since $\alpha_1(\omega)$ contribution dominates the whole intensity
in the $\sigma$-$\pi'$ channel,
we plot the $\psi$ dependence of the pure
$\alpha_1(\omega)$ in this channel (Fig. \ref{Fig.azim} (b)).
Experimental data in Fig. \ref{Fig.azim} (a) are 
those of U in UO$_{2}$ and 
those of Np in U$_{0.75}$Np$_{0.25}$O$_{2}$.\cite{Wilkins04,Wilkins05}
The overall agreement between the experimental data and the 
theoretical curve is interpreted as one of the evidences of the
realization of the triple-\textbf{k} structure in these materials
as the experimentalists have suggested.\cite{Wilkins04,Wilkins05}
Notice that no $\psi$ dependence has been detected
in the $\sigma$-$\pi'$ channel around U and Np $M_4$ edges 
in U$_{0.75}$Np$_{0.25}$O$_{2}$.\cite{Wilkins04,com.2}
Our calculation, however, reveals
that this channel should possess the $\psi$ dependence although 
the dependence is mild around $\psi=-\pi/2 \sim 3\pi/4$
as shown in Fig. \ref{Fig.azim} (b).
The detailed experimental information, in particular, 
around $\psi=\pi$ will clarify this discrepancy. 
Note that, in the $E$1 transition,
quadrupole profile $\alpha^{(2)}(\omega)$
alone has contribution, if any,
on the spectrum in the $sigma$-$\sigma'$ channel.
We have confirmed that
the $\psi$ dependence (not shown) in this channel
expected from the longitudinal triple-{\textbf k} AFQ phase is
completely different from that of the observed one.
Therefore,
it excludes a possibility of any types
of the longitudinal triple-{\textbf k} orderings, such as
the AFM, AFQ and AFO phases.

\section{\label{sec:4}Summary}
We have utilized the useful expression of the RXS amplitude
derived in Ref. 5 to investigate the RXS spectra
at U and Np $M_{4}$ edges in UO$_{2}$ and
U$_{0.75}$Np$_{0.25}$O$_{2}$,
assuming the transverse triple-\textbf{k} AFM ordering phase is
substantiated in these materials. 
Our analysis has shown the semi-quantitative agreement with 
the experimental data, meaning that the spectra in the $\sigma$-$\sigma'$
and $\sigma$-$\pi'$ channels characterize 
the quadrupole ($|\alpha_2(\omega)|^2$)
and dipole ($|\alpha_1(\omega)|^2$)
natures, respectively, as the experimentalists had suggested.
A discrepancy of the azimuthal angle dependence
in the $\sigma$-$\pi'$ channel between the experiments and our theory
should be addressed on the future study.

This work was partially supported by a Grant-in-Aid for Scientific Research 
from the Ministry of Education, Science, Sports and Culture, Japan.


\begin{thebibliography}{99}
\bibitem{com.2}
   As for U site in UO$_{2}$ in Ref. 11, unfortunately, 
   we cannot find any description about the $\psi$ dependence 
   in the $\sigma$-$\pi'$ channel.
\end{thebibliography}
\end{document}